# Effects of Satellite Body on Magnetic Plasma Orbit Control for Nanosatellites

Ryo KAMEYAMA*, Rei KAWASHIMA*, and Takaya INAMORI**

* Department of Electrical Engineering, Shibaura Institute of Technology

** Department of Aerospace Engineering, Nagoya University

**Abstract**

The magnetic plasma orbit control (MPOC) has been proposed for micro and nanosatellites in the sun-synchronous orbits (SSO) in the low earth orbit (LEO). This method utilizes the plasma drag force generated by the interaction between space plasma and the magnetic field surrounding magnetic torquers (MTQs). In this study, the effects of a finite satellite body on high-potential area generation are investigated by using a plasma flow simulation based on the fully kinetic model. The simulation results show that the predicted high-potential region shrinks due to the finite satellite body because the positive charges of stagnated ions in front of the satellite are absorbed into the satellite surface. In addition, simply applying a bias voltage at the front surface is ineffective in expanding the high-potential region. Specifically, applying a positive bias at the front surface resulted in the accumulation of electrons within the satellite enclosure, causing the floating potential of the satellite to drop according to the applied voltage.

**Keywords:** Nanosatellites, Magnetic torquer, Space plasma, Post mission disposal, Particle simulation.

## 1. Introduction

The number of nano- and micro-satellites launched into low earth orbit (LEO) has significantly increased. As a result, space debris due to these small satellites in the LEO has become a serious issue. Nano- and micro-satellites should be equipped with a post-mission disposal method for sustainable use of the LEO. However, due to constraints in weight and space, equipping small satellites with a propulsion system is still challenging. Thus, a deorbit method suitable for nano- and micro-satellites is necessary.

The magnetic plasma orbit control (MPOC) has been proposed as a method for deorbiting nano-satellites [2]. **Figure 1** shows the concept of the MPOC. When a small satellite moves through the ionosphere at orbital speed (~8,000 m/s), electrons, which have less mass than ions, are bound to the magnetic field generated by the satellite. On the other hand, $O^+$ ions have a larger Larmor radius than electrons, and the charge separation between the ions and electrons induces an electric field. This induced electric field repels the ions in front of the satellite due to the Coulomb force, generating a drag force on the satellite. The advantage of MPOC is that the magnetic field is generated by a magnetic torquer (MTQ) that is originally used for satellite attitude control. The MPOC is expected to be used for the post-mission disposal as well as relative orbit control in a formation flight.

The mechanism of drag force generation in MPOC is compared with that of a magnetic sail [4] in **Table 1**. The magnetic sail uses the solar wind in the interplanetary orbit and forms a large magnetosphere structure by using a superconducting magnet. The scale of MPOC is as small as 1.0 m because the satellite onboard MTQ generates the magnetic field. In this case, the satellite body may influence the plasma structure, and one needs to consider the finite satellite size effects in predicting the MPOC performance. In fact, the satellite body has been neglected in the numerical studies of small-scale magnetic sail [5] and MPOC [6], and knowledge on this point is lacking.

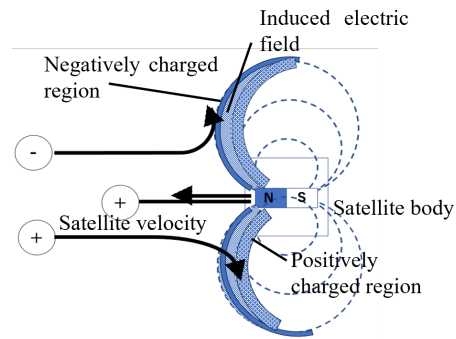

**Fig. 1** Concept of magnetic plasma orbit control.

**Table 1** Comparison between Magnetic sail and Magnetic Plasma Orbit Control (MPOC).

|  | Magnetic sail | MPOC |
|---|---|---|
| Orbital Plasma | Solar wind ($H^+$) | Low earth orbit ($O^+$) |
| Plasma structure scale | $10^5$ m | 1.0 m |

The objective of this study is to numerically predict the effect of the satellite body on the charged particle flow during the operation of the MPOC. We conducted a two-dimensional (2D) plasma simulation in the cases with and without the satellite body. Further, expanding the high-potential region effectively repels the incoming ions. Therefore, we also investigated a method of applying a bias potential at the front of the satellite to assist in forming the high-potential.

## 2. Basics of the Electrostatic Fully Kinetic Model
### 2.1 Plasma Flow
To describe the flow of magnetized plasma, we employ the fully kinetic particle model [5]. This involves a 2D three-velocity particle-in-cell (PIC) method to calculate the motions of ions and electrons. The full-PIC model is essentially consistent with the one used in the previous work [6]. The equation of motion for ions and electrons is solved, respectively, with the electromagnetic force. The present model uses the artificial electron mass model to relax the Courant restriction [6]. Specifically, the electron mass is multiplied by 200 with the artificially increased magnetic field to maintain the Larmor radius.

### 2.2 Electromagnetic Field
In this study, the space potential is obtained through Gauss's law. The 2D Poisson's equation is

$$\nabla \cdot (-\varepsilon \nabla \phi) = e(n_i - n_e), \quad (1)$$

where $\varepsilon$, $\phi$, $n_i$, $n_e$ are the permittivity, space potential, ion number density, and electron number density, respectively. The $\varepsilon$ is equal to the vacuum permittivity in the plasma while it has different values inside the conductor and insulator inside the satellite body. The Poisson's equation is solved by using a second-order scheme. Dirichlet conditions are applied to the Poisson's equation at each boundary surface.

## 3. Simulation Model and Calculation Condition
### 3.1 Satellite Body Model
In this simulation, a finite satellite enclosure is placed at the MTQ position. The satellite body is assumed to be conductive material in which the charge and potential are uniformly distributed. This case corresponds to **Fig. 2(a)**. To make the potential uniform inside the enclosure, the permittivity in Eq. (1) is set to a large value to make the potential gradients small. This setting is consistent with Gauss's law of electrical flux conservation on the surface. The ions and electrons entering the enclosure area are deleted after exchanging their charges with the satellite body.

In case **Fig. 2(b)**, the satellite front surface is isolated by a dielectric material of fluorinated polymer film. In the simulation, a bias voltage is applied at the front surface against the satellite body while the total system is floating. The front surface and satellite body form a capacitor, and the bias voltage is related to the charge separation. The capacitance $C_1$ and charge $Q$ are respectively calculated as follows

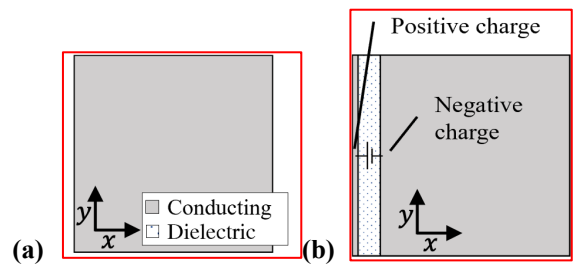

**Fig. 2** Satellite body model. (a) The body is all conductor and (b) the front surface is positively biased with a thin layer of insulator.

$$C_1 = \varepsilon \frac{S}{d}, \quad Q = C_1 V_b, \quad (2)$$

where $S$, $d$ $V_b$ represent the surface area and thickness of the dielectric thin film, respectively, and $V_b$ is the bias voltage. In this simulation, the charge $+Q$ and $-Q$ are respectively deposited on the front surface and satellite body to apply the bias voltage.

To account for the difference in size between the dielectric film and the simulation cells, the film ($d = 5.0 \times 10^{-4}$ m) is stretched to match the cell, and the dielectric constant is adjusted to maintain the same capacitance as before stretching. The dielectric constant is analyzed separately in the x- and y-directions as follows:

$$\varepsilon_x = \varepsilon \frac{\Delta x}{d}, \quad \varepsilon_y = 1.0 \times 10^6, \quad (3)$$

where the $\varepsilon_y$ is set high enough to work as a conductor along the satellite surface.

### 3.2 Simulation Model
**Table 2** lists the assumed parameters for the simulation of the magnetic plasma deorbit in LEO. The simulation model for this study is shown in **Fig. 3.** The properties of plasma are calculated through the International Reference Ionosphere (IRI) 2016 model [7] as functions of altitude and time. In this simulation, a small satellite operating in a polar orbit at an altitude of $h = 570$ km was used as the calculation condition, referring to data from the small satellite MAGNARO [8].

The ion species was set to be oxygen ion $O^+$, which is more than 90 percent in the ionosphere of LEO. The MPOC generates a large drag force as the orbital plasma density is high. In this study, the satellite is assumed to have a sun-synchronous orbit (SSO). Along the SSO, the plasma density is high when the satellite is around the equator. Thus, this study assumes that the MPOC is used around the equator, and the relatively high plasma density at $h = 570$ km is assumed. The plasma density and satellite velocity determine the inflow condition at the left boundary in Fig. 3.

**Table 2** Parameters of nominal calculation condition.

| Parameter | Value |
| --- | --- |
| Ion number density, m$^{-3}$ | $2.7 \times 10^{11}$ |
| Ion species, - | O$^+$ |
| Ion temperature, K | 1500 |
| Electron temperature, K | 2500 |
| Satellite velocity, m/s | 8000 |
| Magnetic moment, Am$^2$ | 10 |
| MTQ orientation, - | Parallel to plasma flow |
| Satellite body size | 0.14 m × 0.14 m |

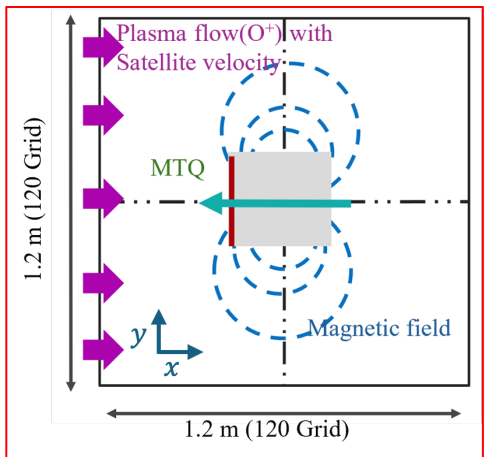

**Fig. 3** Calculation domain.

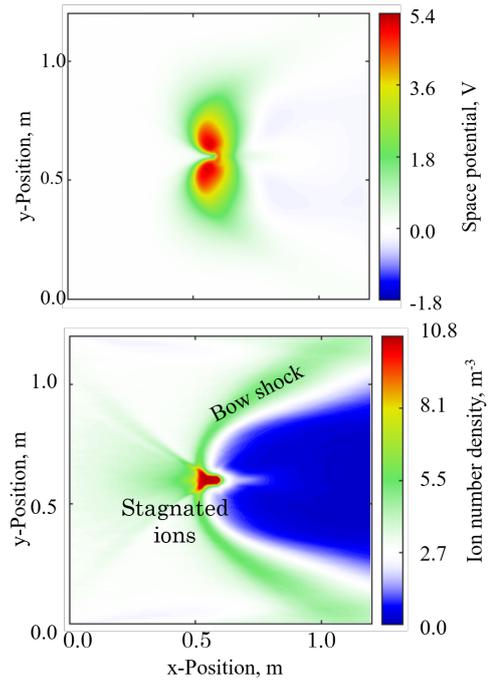

**(a)** No satellite enclosure case

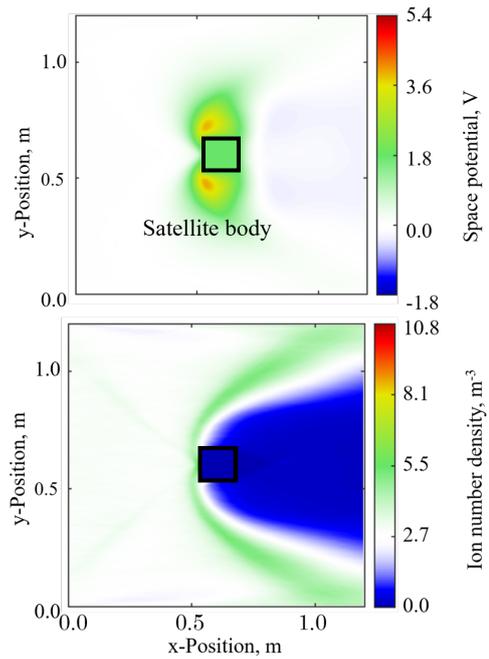

**(b)** Place satellite enclosure case

**Fig. 4** Distribution of space potential and ion number density in the cases with and without the satellite enclosure.

## 4. Results and Discussion
*4.1 Effect of placing the satellite enclosure*

**Figure 4(a)** shows the distributions of space potential and ion number density when the satellite enclosure is not included in the calculation domain. A potential structure decelerating the incoming ions is formed with a peak value of 5.4 V. The kinetic energy of incoming ions corresponds to the potential of 5.3 V, which indicates that the incoming ions can be braked by the potential structure surrounding the MTQ. The ion number density distribution exhibits the stagnated ions in front of the MTQ. A similar plasma structure was observed in the laboratory experiment, which qualitatively validated the present simulation results [9].

**Figure 4(b)** presents the simulation results with the finite satellite enclosure placed at the MTQ position as shown in Fig. 2 (a). In this case, the maximum potential is 4.3 V, and the high-potential region that bounces the ions back shrinks. This potential is insufficient to repel the incoming ions of the kinetic energy of 5.3 V. The reason for the shrunk high-potential region is the absence of stagnated ions in front of the satellite. The ions collide with the satellite body, exchanging their charge with the surface. Because the positively charged region in front of the satellite diminishes, the potential structure formed by the MTQ becomes less steep.

It has been found that the finite satellite body has negative effects on forming the high-potential region and reflecting the incoming ions. The simulated steady plasma drag forces in the cases with and without the satellite enclosure are 70.1 nN and 238.8 nN, respectively. The simulated drag forces can be used for a deorbit scenario where drag force is generated in high-plasma density regions. A fast satellite deorbit may be achieved if one uses an intensive drag force generation to make the perigee altitude as low as possible. In the subsequent study, we will investigate a deorbit scenario in which a large drag force is generated in the high plasma density regions.

To increase the drag force in the case with the satellite enclosure, the maximum potential must be greater than 5.3 V. One method to increase the maximum potential is to use a larger magnetic moment. However, the magnetic moment of 10 Am$^2$ is already large for the attitude control of nanosatellites. Thus, we investigate a method to enhance the drag force, without increasing the magnetic moment.

*4.2 Effect of applied potential*

**Figure 5** indicates the distribution of space potential when a bias voltage of 3 V is applied as shown in Fig. 2 (b). Even with the applied potential, the potential structure in the plasma is similar to that of the conducting satellite body case in Fig. 4(b). **Figure 6** shows the space potential distribution along the horizontal line at $y = 0.6$ m in the cases of different bias voltages. This result indicates that the shape of the high potential region remains unchanged even when the applied bias voltage at the front surface increases. This is because electrons around the satellite are accumulated within the satellite enclosure, decreasing the satellite's floating potential. It was found that applying the bias voltage is ineffective in enhancing the high potential region in the MPOC. Note that the slight potential shifts at the left and right boundaries are supposed to stem from an imperfect inflow condition for electron particles. Its effects on the potential structure around the satellite is minor.

**5. Conclusions**

We performed a 2D PIC simulation to investigate the effect of the satellite body on MPOC for nanosatellites. The major findings are summarized as follows.
(1) The high-potential region shrinks due to the finite satellite body because the positive charges of stagnated ions in front of the satellite are absorbed into the satellite front surface.
(2) Simply applying a bias voltage at the front surface is ineffective in expanding the high-potential region. Applying a positive bias at the front surface resulted in the accumulation of electrons within the satellite enclosure, and floating potential of the satellite dropped according to the applied voltage.

**Acknowledgment**

This work was supported by JSPS KAKENHI Grant Number JP21H01531.

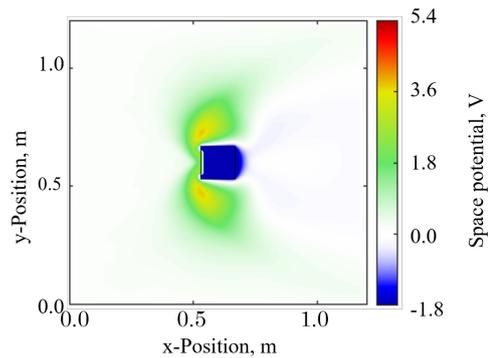

**Fig. 5** Distribution of space potential when 3 V is applied at the front surface of satellite enclosure.

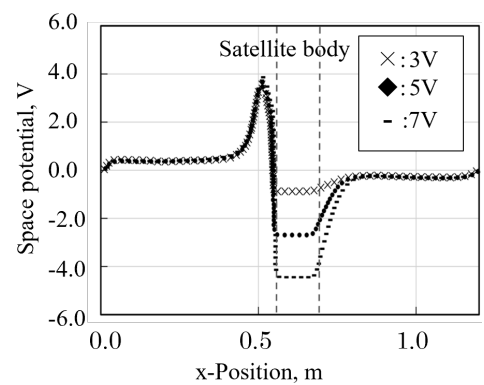

**Fig. 6** Space potential across the satellite in the x-axis direction when the bias voltage increases.

**References**
1) T. Inamori et al., "Magnetic plasma deorbit system for nano- and micro-satellites using magnetic torquer interference with space plasma in low Earth orbit," *Acta Astro.*, **112** (2015), 192–199.
2) I. Funaki et al., "Laboratory Experiment of Plasma Flow Around Magnetic Sail," *Astrophys. Space. Sci.*, **307** (2007), 63–68.
3) Y. Ashida et al., "Thrust Evaluation of Small-Scale Magnetic Sail Spacecraft by Three-Dimensional Particle-in-Cell Simulation," *J. Propul. Power*, **30**-1 (2014), 186–196.
4) R. Kawashima et al., "Particle Simulation of Plasma Drag Force Generation in the Magnetic Plasma Deorbit," *J. Spacecr. Rockets*, **55**-5 (2018), 1083–1097.
5) D. Bilitza et al., "International Reference Ionosphere 1990," *NASA Technical Memorandum*, NASA-TM-105055 (1990), 0–84.
6) T. Inamori et al., "In Orbit Demonstration of Propellant-Less Formation Flight with Momentum Exchange of Jointed Multiple CubeSats in the MAGNARO Mission," 36th Annual Small Satellite Conference, Utah, 2022.
7) V. A. Shuvalov, et al., "Drag on a spacecraft produced by the interaction of its magnetic field with T the Earth's ionosphere. Physical modelling," *Acta Astronautica*, **166** (2020), 41−51.